\documentclass{article}

\usepackage{arxiv}

\usepackage[utf8]{inputenc} % allow utf-8 input
\usepackage[T1]{fontenc}    % use 8-bit T1 fonts
\usepackage{hyperref}       % hyperlinks
\usepackage{url}            % simple URL typesetting
\usepackage{booktabs}       % professional-quality tables
\usepackage{amsfonts}       % blackboard math symbols
\usepackage{nicefrac}       % compact symbols for 1/2, etc.
\usepackage{microtype}      % microtypography
\usepackage{lipsum}         % Can be removed after putting your text content
\usepackage{graphicx}
\usepackage{doi}
\usepackage{graphicx}
\usepackage{amssymb}  
\usepackage{amsmath}
\usepackage{pgfplots}
\usepackage{cleveref}       % smart cross-referencing

\title{Time Optimization of Constrained Control\\ for a Thermoelectric Solid System\\ with a Peltier Element \thanks{The study has been done under financial support of
	the Russian Science Foundation (grant 21-11-00151).}
}

\author{ 
	\href{https://orcid.org/0000-0001-7210-5395}{\includegraphics[scale=0.06]{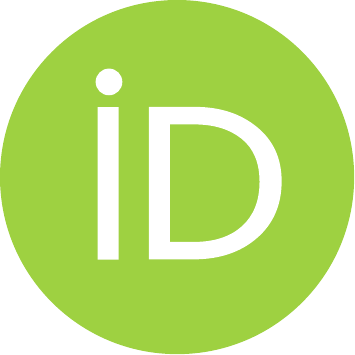}\hspace{1mm}Alexander Gavrikov} \\
	Department of Mathematics\\
	Penn State University\\
	State College, USA \\
	\texttt{avg6113@psu.edu} \\
	Ishlinsky Institute for Problems in Mechanics\\ of the Russian Academy of Sciences (IPMech RAS)\\
	Moscow, Russia \\	
		\texttt{gavrikov@ipmnet.ru} \\	
	\And
\href{https://orcid.org/ 0000-0001-6526-6246}{\includegraphics[scale=0.06]{orcid.pdf}\hspace{1mm}Georgy Kostin}
 \\	Ishlinsky Institute for Problems in Mechanics\\ of the Russian Academy of Sciences (IPMech RAS)\\
	Moscow, Russia \\	
		\texttt{kostin@ipmnet.ru} \\
}

\renewcommand{\shorttitle}{Time Optimization of Constrained Control}

\begin{document}
\def\ds{\displaystyle}
\maketitle

\begin{abstract}
A solid system consisting of two heat conducting cylinders with a thermoelectric converter (Peltier element) between them is considered. A nonlinear model, which was previously verified by authors, is used to design a constrained control law that allows us to achieve a steady-state distribution of the temperature in one of the cylinders in much less time than the characteristic time of transient processes. The initial-boundary value problem is exactly linearized over temperature by means of feedback linearization. Although the resulting system is nonlinear in a control function, it is possible to construct a finite-dimensional approximation  based on analytical solution of the corresponding eigenproblem for a constant control signal. The time-optimal control problem  is studied numerically by using this eigenfunction decomposition.  To construct admissible control laws, an auxiliary unconstrained optimization problem is introduced. Its cost functional  represents a weighted sum of temperature deviation from the desired zero distribution and a penalty for violating an electric power constraint.  The control time interval is split into several parts, and on each subinterval the control signal is taken constant. The optimal piecewise constant feedforward control is found numerically by applying the gradient descent method. We analyze the proposed control law with respect to the shortest admissible time of the process.
\end{abstract}

% keywords can be removed

\section{Introduction}

At present, solid-state thermoelectric converters, including Peltier elements (PEs), are increasingly used in technical applications. Despite moderate power of such devices, they have a number of significant advantages, such as compactness, resistance to external disturbances, accuracy and fast response rate of control signals. Applying thermoelectric converters as actuators and sensors makes it possible to create efficient heat transfer systems and control various technological processes to achieve and maintain certain temperature regimes in solids.  The examples of such application, including systems utilizing  renewable energy sources, can be found in farming \cite{Tikhomirov:ER:2020}, machining \cite{Mironova:IEEETIA:2020}, energy generation  \cite{Fan:ER:2020}, air conditioning \cite{Moria:ATE:2021}, water heating \cite{Kwan:ECM:2019} and cooling \cite{Tijani:RE:2018}. For the control of such systems, a wide variety of methods is used from bang-bang type control \cite{Badescu:JNET:2021} to classical PI controllers as well as neural network control models \cite{Honc:ISASE:2019}.

To construct accurate control laws, it is necessary to use coupled nonlinear thermoelectric models that take into account  effects of electric energy recuperation, Joule heat losses, and heat exchange with the environment. Given  limitations imposed on the power consumption of a PE, it is important to evaluate both  reachability ranges for admissible input signals  and the minimal time in which the system can be transfered to the desired state. Even for relatively simple systems, a direct solution of the time-optimal control problem seems to be a rather difficult goal. Therefore, one can use different approaches to numerically estimate the minimum of control time. In this paper, we apply the gradient descent method for this purpose. For a system consisting of two metal cylinders and a Peltier disk-like element separating them, we try to reach a small neighborhood of the desired terminal state, while the constraints on the electric current are taken into account utilizing a penalty term added to the cost functional.

\section{Model of a Thermoelectric Converter}
\begin{figure}
\begin{center}
\includegraphics[width=6.3cm]{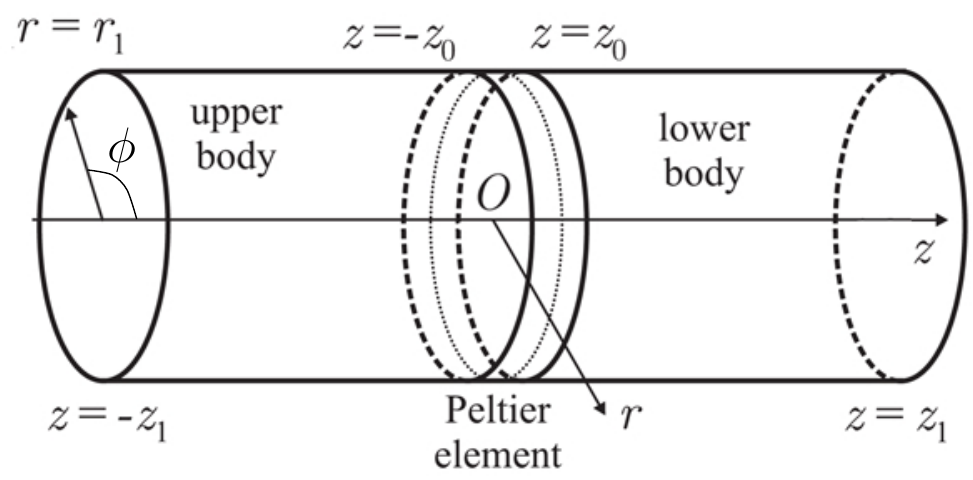}   
\caption{Schematic representation of the experimental setup.} 
\label{fig:setup}
\end{center}
\end{figure}

A full controlled plant may contain many actuated elements consisting of parts with complex geometry and inhomogeneous parameters, what hinders a study of limiting behavior. For this reason, as a simple example we consider a thermoelectric system consisting of two identical heat conducting cylinders and a thin circular PE between them, see Fig.~\ref{fig:setup}. The length of the cylinders and the PE are $z_1-z_0$ and $2z_0,$ respectively, where $z_0\ll z_1.$ Their radii are equal to $r_1.$ A corresponding experimental setup was constructed at the Chair of Mechatronics of the University of Rostock, Germany 
\cite{Gavrikov:MMAR:2019-1}, \cite{Knyazkov:MMAR:2019}.

Under assumption that the PE characteristics are constant and heat transfer in the PE occurs  only in one direction, a nonlinear model of  thermoelectric processes was proposed and validated in \cite{Gavrikov:MMAR:2019-1},  \cite{Knyazkov:MMAR:2019}, \cite{Kostin:IFACPOL:2019},  \cite{Gavrikov:STAB:2020}. The governing equations of this model  were simplified by introducing a feedback linearization control in \cite{Gavrikov:MMAR:2019-2}. The resulting initial-boundary value problem (IBVP) has the following form in cylindrical coordinates  $\mathbf{x}=(r,\phi,z)$
\begin{equation}\label{IBVP-Cylinders with Feedback Linearization}
\begin{array}{c}
c_p \rho_p\dot\theta= \lambda_p \theta''_{zz} + \frac{u^2}{R|\mathcal{V}_p|},\quad \mathbf{x}\in \mathcal{V}_p,\\
c_a \rho_a\dot\theta= \lambda_a \Delta \theta ,\quad \mathbf{x}\in \mathcal{V}_k,\quad k=1,2,\\
\theta'_z|_{|z|=z_1}=0,\quad
\alpha\theta + \lambda_a\theta'_r|_{r=r_1, |z|>z_0}=\alpha\theta_A,\\
\theta|_{|z|= z_0\pm 0}=\theta|_{|z|= z_0\mp 0},\\
\left. - \lambda_a\theta'_z\right|_{|z|=z_0+0} = \left[(\theta+\theta^0)\frac{Su}{R|\mathcal{A}_p|}- \lambda_p\theta'_z\right]_{|z|=z_0-0},\\
\theta(0,\mathbf{x})=\Theta(\mathbf{x}),\\
u(t)=
\left\{
\begin{array}{ll}
u_0(t)-u_-& \mathrm{for}\quad u_0<u_-\\
0& \mathrm{for}\quad u_-\leq u_0\leq u_+\\
u_0(t)-u_+& \mathrm{for}\quad u_0>u_+\\
\end{array}.
\right.
\end{array}
\end{equation}
%with the initial condition $\theta(0,\mathbf{x})=\Theta(\mathbf{x}).$
Here, $c_{p,a}$ are specific heat capacities, $\rho_{p,a}$ are densities, $\lambda_{p,a}$ are thermal conductivities for the PE ($p$) and the cylinders ($a$), $\alpha$ is the heat transfer coefficient,  $\theta_A$ and $\theta^0$ are the constant ambient and reference temperatures, respectively, $|\mathcal{A}_p|=\pi r_1^2,$ $S$ is the Seebeck coefficient, and $R$ is  the ohmic resistance of the PE.  The cylinders and the PE occupy domains $\mathcal{V}_k=I_r\times I_\phi\times I_k$  $(k=1,2)$ and   $\mathcal{V}_p=I_r\times I_\phi\times I_p$,  respectively, where $I_r=[0,r_1],$ $I_\phi=[0,2\pi],$ $I_p=[-z_0,z_0],$ $I_1=[z_0,z_1],$ $I_2=[-z_1,-z_0].$ The temperature distribution $\theta(t,\mathbf{x})$ and the ambient temperature are measured relatively to the reference  temperature $\theta^0$.
Although we suppose that $\theta_A$ is constant, the proposed approached can be extended to the ambient temperature varying in time \cite{Gavrikov:IFAC:2020}. For definiteness, it is assumed that the thermoelectric system (\ref{IBVP-Cylinders with Feedback Linearization}) is initially in a steady state: $\Theta(\mathbf{x})=\theta_{st},$ such that the cylinder $\mathcal{V}_1$ has average temperature $\theta_{av}$, and $u_{st}$ is a constant voltage that yields the temperature distribution $\theta_{st}$.

 The control function $u(t)$ is defined through the input voltage $u_0(t)$ and thresholds  $u_{\pm}$. The total control voltage supplied to the PE is expressed as 
\begin{equation}\label{full control}
U=u_0+S[\tilde{\theta}].
\end{equation}
Here,  $[\tilde{\theta}]$ denotes the jump of the average temperature on the top and bottom sides of the PE. Although the voltage $S[\tilde{\theta}]$ provides the  linearization  w.r.t. the temperature, see \cite{Gavrikov:MMAR:2019-2}, the resulting IBVP (\ref{IBVP-Cylinders with Feedback Linearization}) is still nonlinear in $u$. Therefore, an explicit design of optimal control is hindered. In what follows, we propose a semi-analytical approach to solution of an optimal control problem (OCP) for~(\ref{IBVP-Cylinders with Feedback Linearization}) based on eigenfunction decomposition of $\theta$ for constant values of $u$.

\section{Optimal Control Problem}
  
Previously, we considered an OCP for the IBVP (\ref{IBVP-Cylinders with Feedback Linearization}), where the goal was to achieve a steady-state distribution of the temperature in one of the cylinders. Although both cylinders are actuated by the PE in such a problem, the second cylinder $\mathcal{V}_2$ serves only as a ``heat sink'', while the first one, $\mathcal{V}_1$, represents a controlled part of a plant in which a stable working regime has to be achieved. To this end, we utilized a piecewise control function \cite{Gavrikov:MMAR:2019-2} and took into account disturbances of the ambient temperature \cite{Gavrikov:IFAC:2020}. Recently, we optimized a constrained piecewise control signal \cite{Gavrikov:APM:2022}.  

\begin{figure}
\begin{center}
\includegraphics[width=5.6cm]{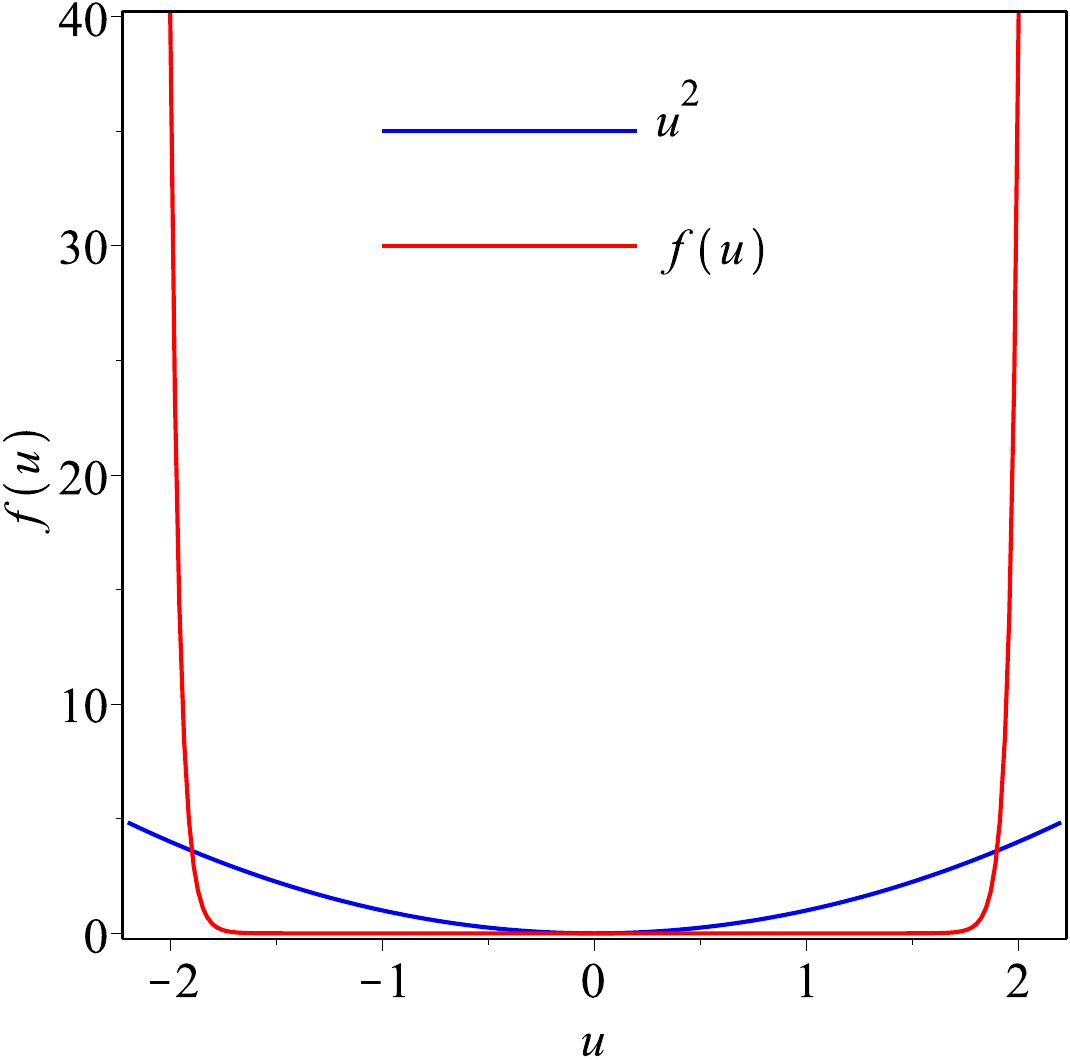} 
\caption{Penalty function $f(u)$ vs $u^2$.} 
\label{fig:penalty function}
\end{center}
\end{figure}	 

\subsection{Control problem: general statement}

In this work, we consider a time-optimal control problem.  We suppose that the goal is to return the actuated cylinder to the zero state in a shortest time $T:$
\begin{equation}\label{OCP general}
T \to \min_u\text{ subject to } \frac{\|\theta(T,\mathbf{x})\|_{L_2(\mathcal{V}_1)}}{\|\theta(0,\mathbf{x})\|_{L_2(\mathcal{V}_1)}}\leqslant\varepsilon
\end{equation}
as well as  the IBVP (\ref{IBVP-Cylinders with Feedback Linearization}) and constraints on $u$:
\begin{equation}\label{control constraints}
|u|\leq |u_{st}|.
\end{equation}
That is, we suppose that the bounds  on $u$ are defined by a control signal $u_{st}$ corresponding to the initial steady state $\theta_{st}.$

\subsection{Relaxed control problem}

Direct solution of (\ref{OCP general}), (\ref{control constraints}) is hindered due to its nonlinearity and uncertainty of existence of global or even local minimum. To deal with these difficulties, we consider a relaxed version of (\ref{OCP general}), (\ref{control constraints}) instead of finding exact solution. We find admissible control laws such that $\frac{\|\theta(T,\mathbf{x})\||_{L_2(\mathcal{V}_1)}}{\|\theta(0,\mathbf{x})\||_{L_2(\mathcal{V}_1)}}\leqslant \varepsilon$ and (\ref{IBVP-Cylinders with Feedback Linearization}), (\ref{control constraints}) are satisfied. That allows for estimating from above the minimal control time $T_{min}$. To design numerically admissible control laws, we consider an unconstrained optimization problem keeping only (\ref{IBVP-Cylinders with Feedback Linearization}) as a constraint.
To this end, we minimize
 %, that is to solve \eqref{OCP general}, \eqref{control constraints}, 
%the auxiliary unconstrained optimization problem is considered in terms of 
the cost function
\begin{equation}\label{Full Cost Functional}
\begin{array}{c}
F \to \min\limits_u,\quad
F=10^\gamma F_{d} +  F_p, \quad %\gamma>0,\\[1.5ex]
F_d=c_\theta \|\theta(T,\mathbf{x})\|^2_{L_2(\mathcal{V}_1)},\\[1.5ex] \displaystyle c_\theta=({|\mathcal{V}_1|\theta_{av}^2})^{-1}{2\pi\int_0^{r_1} rJ_0^2(\mu_{0,0}r/r_1)dr},
\end{array}
\end{equation}
subject to the IBVP (\ref{IBVP-Cylinders with Feedback Linearization}). In (\ref{Full Cost Functional}),  $J_0$ is the Bessel function of zeroth order of the first kind, $\mu_{0,0}$ is a root of an equation presented in the next section. Note that the $L_2$-norm in (\ref{Full Cost Functional}) is evaluated only over  $\mathcal{V}_1.$ That is, we minimize  the temperature only in the controlled cylinder $\mathcal{V}_1$, while the actuated cylinder $\mathcal{V}_2$ may have any temperature  yielded by $u$.

The functional $F_d$ in (\ref{Full Cost Functional}) represents the norm of the deviation of the temperature distribution from the zero state in the controlled cylinder $\mathcal{V}_1$ at time $T$, while
\begin{equation}\label{penalty functional}
F_p=c_u\int_0^T f(u) dt,\quad c_u=(\rho_a c_a R |\mathcal{V}_1|\theta_{av})^{-1},
\end{equation}
is a penalty term. 

The  penalty functional $F_p$ was  proposed in \cite{Gavrikov:APM:2022} and characterized as follows: 
\begin{equation}\label{f(u) conditions}
\begin{array}{c}
f(u)\sim o(u^2) \text{ for } |u|< \varepsilon, \quad \varepsilon\ll u_{st},\\
f(u)\ll u^2 \quad\text{for}\quad |u|<u_{st}-\varepsilon,\\
f(u)\gg u^2 \quad\text{if}\quad |u|\geq u_{st}.
\end{array}
\end{equation}
Therefore, $F_p$ is small enough if the control function satisfies the constraints and is large otherwise.
We define the penalty function according to
\begin{equation}\label{penalty function}
f(u)=c_1(e^{c_2u^2}-c_2u^2-1).
\end{equation}
The constants $c_1, c_2$ must be chosen such that the conditions (\ref{f(u) conditions}) are satisfied. See  Fig.~\ref{fig:penalty function}, where these constants are as in the numerical example presented further.

To estimate the minimal control time $T_{min}$ in (\ref{OCP general}), we solve the problem (\ref{Full Cost Functional}) for an admissible solution via the gradient descend method for $T$ in some range $[T_0,T_1]$ choosing as $T_{min}$ such a value that the constraints  (\ref{control constraints}) are satisfied.

\section{Finite Dimensional Approximation}\label{Finite Dimensional Approximation}
To simplify the studied control problem, we consider a finite dimensional approximation of the IBVP (\ref{IBVP-Cylinders with Feedback Linearization}). Suppose that $u(t)$ is constant: $u(t)=u^{(i)}$. Then a solution to (\ref{IBVP-Cylinders with Feedback Linearization}) can be represented  as the series
\begin{equation}\label{Ansatz Functions}
\begin{array}{c}
\theta=\sum_{n,m,k}e^{\nu_{n,m,k} t}\Xi_{n,m,k}(r,\phi,z),\\[1.5ex]
\Xi_{n,m,k}(r,\phi,z)=J_n(\mu_{n,m} r/r_1) \cos(n \phi) \psi_k (z),%,\, n=0,1,\ldots,
\end{array}
\end{equation}
where $\Xi$ are orthogonal eigenfunctions of the IBVP (\ref{IBVP-Cylinders with Feedback Linearization}).
In (\ref{Ansatz Functions}),  $J_n$ are Bessel functions of the first kind of order $n$, $\mu=\mu_{n,m}$ are roots of the  equation
\begin{equation}\label{Boundary Condition Involving Bessel Functions}
\alpha J_n(\mu) + \frac{\lambda_a}{r_1}(nJ_n(\mu)-\mu J_{n+1}(\mu))=0,
\end{equation}
which corresponds to the boundary condition on the lateral  surface of the cylinders.
The quantities $$\nu=\tau^{-1}=-\frac{\lambda_a(\mu^2 + \xi^2)}{c_a\rho_a r_1^2}$$ are the inversed characteristic times of decay $\tau=\tau_{n,m,k}$ of each  mode, and $\xi=\xi_k$ are eigenvalues corresponding to the eigenfunctions in the $z$-direction $\psi(z)=\psi_k(z)$ \cite{Gavrikov:MMAR:2019-2}. These functions are found explicitly as  piecewise continuous linear combinations of exponential and trigonometric functions.

Analysis of characteristic times   $\tau_{n,m,k}$ allows one to determine the modes  that can be neglected on the time interval $[0,T]$ due to their quick decaying. For the experimental setup considered here, only the four lowest modes have characteristic times   $\tau_{n,m,k}\gtrsim 10$~sec. For the highest modes, $\tau_{n,m,k}\lesssim 5$~sec, see \cite{Gavrikov:MMAR:2019-2}. Therefore, these transient processes can be excluded from consideration if the intervals of constancy of $u$ are much longer. It is worth mentioning that the angular modes ($n>0$) are uncontrollable in this setting. However, they belong to the group of quickly decaying modes and can be neglected.

Hence, the expansion (\ref{Ansatz Functions}) simplifies to 
\begin{equation}\label{four mode expansion}
\theta=\sum_{k=0}^3e^{\nu_{0,0,k} t} J_n(\mu_{0,0} r/r_1) \psi_k (z).
\end{equation}
The expansion of the initial steady state $\Theta(\mathbf{x})=\theta_{st}$ w.r.t. to these eigenfunctions yields a vector of coefficients $\bar\Theta$.
Next, by substituting (\ref{four mode expansion}) into (\ref{IBVP-Cylinders with Feedback Linearization}) and projecting the result onto the eigenfunctions, we obtain the four-dimensional system of ordinary differential equations
\begin{equation}\label{ODE system Feedback Linearization}
\begin{array}{c}
\dot{\bar\theta}= A\bar\theta + 
{G}(u,\theta_A,\theta^0),\quad
\bar\theta(0,z)=\bar\Theta,\\
\bar\theta=(\theta_{0},\ldots,\theta_{3})^T,\quad \bar\Theta=(\Theta_{0},\ldots,\Theta_{3})^T,\\
\displaystyle \Theta_{i}=\int_{\mathcal{V}} \Theta(\bar x)\Xi_i(\mathbf{x})d\mathbf{x},\quad  \Xi_i=J_0\left(\mu_{0,0}r/{r_1}\right)\psi_{0,0,i}.
\end{array}
\end{equation}
Here, $\theta_i=\theta_{0,0,i}$, the matrix $A$ is diagonal consisting of eigenvalues $\nu_{0,0,i}$,  and the vector    $G$ is constant for a constant $u$. Hence, the system (\ref{ODE system Feedback Linearization}) can be solved explicitly for any given initial conditions (ICs) $\bar\Theta$.

The steady state $\theta_{st},$ which we take as ICs in (\ref{IBVP-Cylinders with Feedback Linearization}), is obtained for a given voltage $u_{st}$ in a similar way. Neglecting the rate of temperature change in (\ref{IBVP-Cylinders with Feedback Linearization}), we plug into the resulting system the ansatz function $\theta_{st}(\mathbf{x})=J_0(\mu_{0,0} r/r_1) \psi_{st} (z)$ and find $\psi_{st}$ as a function of $u_{st}$ explicitly  as a piecewise continuous linear combination of cosines and quadratic polynomials.

In its turn, the voltage $u_{st}$ is chosen such that the resulting temperature distribution $\theta_{st}$ has some prescribed average  value $\theta_{av}$ in the actuating cylinder: $\theta_{av}=\frac{1}{|\mathcal{V}_1|}\int_{\mathcal{V}_1} \theta_{st} dV.$ Utilizing  the procedure described above and functions $\psi_{st}(u_{st})$, we find $u_{st}$ explicitly  by solving the minimization problem
\begin{equation}\label{Minimum Condition Steady State Minus Average}
\Big(\frac{1}{|\mathcal{V}_1|}\int\nolimits_{\mathcal{V}_1} \theta_{st} dV - \theta_{av}\Big)^2 \to\min\limits_u.
\end{equation}

\section{Solution of the control problem}\label{Solution of optimal control problem}

\subsection{Auxiliary optimization problem}

\begin{figure}
\begin{center}
\includegraphics[width=0.6\textwidth]{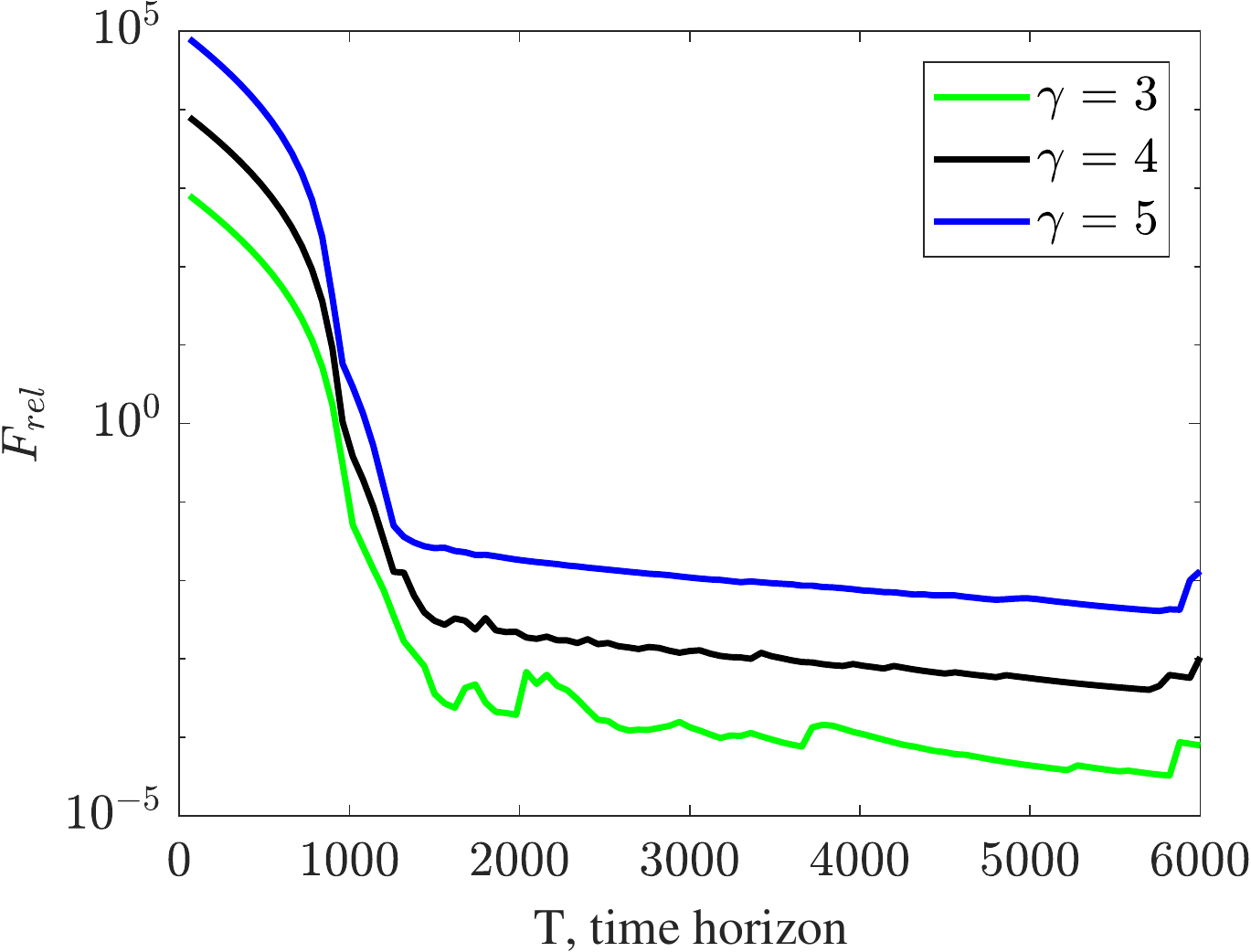}
\caption{The relative value of the cost functional $F$.} 
\label{fig:cost functional}
\end{center}
\end{figure}	

In this section, a combined numerical-analytical  approach to the solution of OCP (\ref{OCP general}) is proposed assuming that the control function $u(t)$ is piecewise constant.  

For a given $T$, the time interval $[0,T]$ is split into 5 equal subintervals $[t_{i},t_{i+1}]$, $t_0=0,$ $t_5=T$, $i=0,\ldots,5$. We suppose that $u(t)$ is constant on each subinterval: $u(t)=u^{(i)}\in\mathbb{R}$ for $t\in[t_i,t_{i+1}].$  For a given $u^{(i)}$, the eigenfunctions $\psi_k^{(i)} (z)$ are found according to Sect.~\ref{Finite Dimensional Approximation}. Next, we solve explicitly (\ref{ODE system Feedback Linearization}) on $[0,t_1]$ for the given initial steady state $\bar\Theta$ and the control function $u(t)=u^{(0)}.$ The resulting distribution $\theta(t_1,\mathbf{x})$, which is a linear combination of eigenfunctions $\psi_k^{(0)} (z)$,  is re-expanded w.r.t. the eigenfunctions $\psi_k^{(1)} (z)$ corresponding to $u^{(1)}$. Taking coefficients of this expansion as ICs $\bar\Theta^{(1)}$, we solve on $[t_1,t_2]$ the system (\ref{ODE system Feedback Linearization}) with a new matrix $A^{(1)}$ determined by eigenvalues $\nu_k^{(1)}$. % eigenfunctions $\psi_k^{(1)} (z)$. 
This yields the temperature $\theta(t_2,\mathbf{x})$. We repeat consequently solution of (\ref{ODE system Feedback Linearization}) and re-expansion of $\theta$ until  the terminal state $\theta(T,\mathbf{x})$ is reached.

For the fixed time horizon $T$, we find piecewise constant $u=(u^{(0)},\ldots,u^{(4)})^T$ such that
\begin{equation}\label{Minimization problem}
F[u,\theta]\to\min_{u(t),t\in[0,T]},
\end{equation}
where $F$ is defined in (\ref{Full Cost Functional}). This problem is solved via the gradient descend. When starting with a vector
\begin{equation}
\bar u^{(0)}=(u^{(0,0)},\ldots,u^{(0,4)}),
\end{equation}
the next value is computed according to
\begin{equation}
\bar u^{(j+1)}=\bar u^{(j)}+\varepsilon \nabla_{\bar u} F,\quad\varepsilon\ll 1.
\end{equation}
The gradient $\nabla_{\bar u} F$ is found numerically via finite differences
\begin{equation}
\frac{F(u^{(j,l)}+\delta u^{(j,l)})-F(u^{(j,l)})}{\delta}, \quad \delta\ll 1,
\end{equation}
but the values of $F(u^{(j,l)})$ and their variation $F(u^{(j,l)}+\delta u^{(j,l)})$  are obtained analytically by using the algorithm presented in Sect.~\ref{Finite Dimensional Approximation}.

\subsection{Estimating the minimal time}

\begin{figure}
\begin{center}
\includegraphics[width=0.6\textwidth]{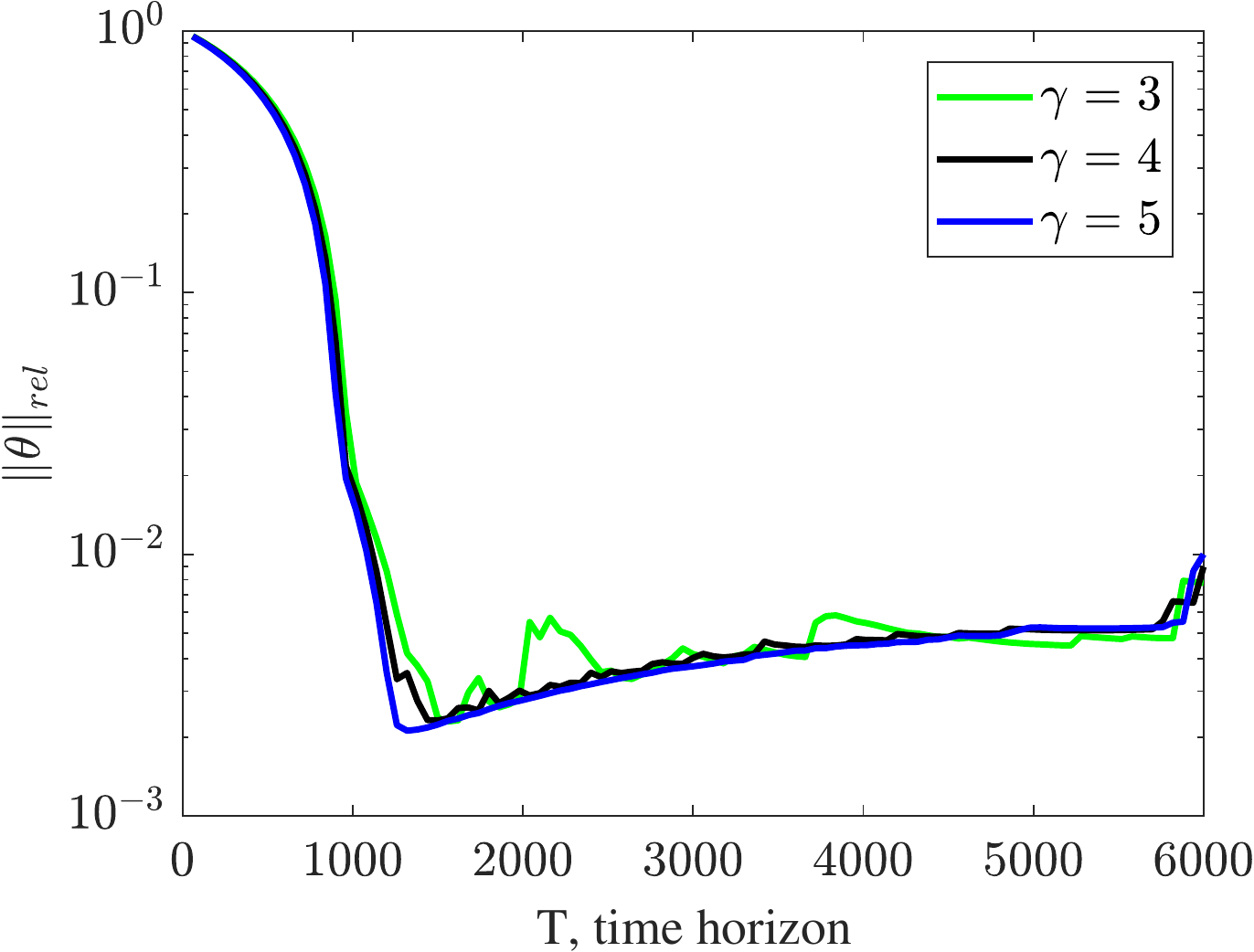}
\caption{The norm of the terminal temperature distribution in the controlled cylinder related to the norm of temperature at $t=T$ due to natural cooling.} 
\label{fig:temperature normalized by cooling}
\end{center}
\end{figure}	

To estimate the minimal time $T_{min}$, we start with $T_0\sim \tau_{0,0,0}$, where $\tau_{0,0,0}$ is the characteristic time of decay of the zeroth eigenmode. For such large $T$, the system loses most of the heat due to natural cooling/heating and the active control is almost not needed. Next, we decrease $T$, $T_j=T_{j-1}-\delta T_0$ and solve the OCP (\ref{Minimization problem}) on the time interval  $[0,T_j].$ The minimal value of $T_j$ such that the solution $u(t)$ of (\ref{Minimization problem}) does not violate the constraints, we take as an estimate from above of $T_{min}$. 

In the next section, we present a numerical example of implementation of approaches described in Sects.~\ref{Finite Dimensional Approximation},~\ref{Solution of optimal control problem}.

\begin{figure}
\begin{center}
\includegraphics[width=0.6\textwidth]{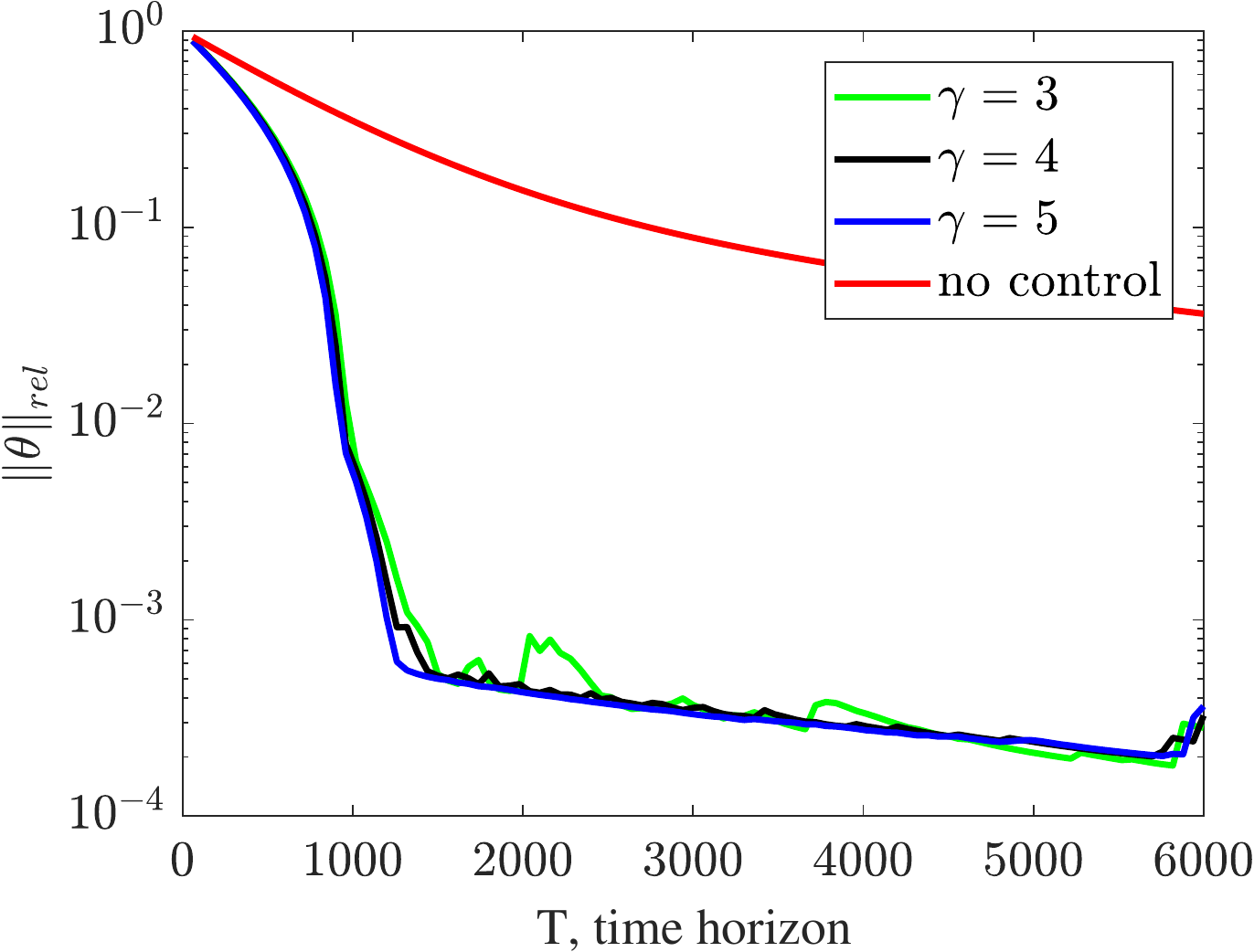}
\caption{The norm of the terminal temperature distribution in the controlled cylinder related to the norm of initial temperature distribution.} 
\label{fig:temperature normalized by initial}
\end{center}
\end{figure}

\section{Numerical results and discussion}
The following physical parameters corresponding to the experimental setup, see \cite{Knyazkov:MMAR:2019}, \cite{Gavrikov:MMAR:2019-1}, are considered: 
$\lambda_a=254.4$~W/m/K, $\lambda_p=0.517$~W/m/K, $\rho_a=2700$~kg/m$^3,$ $\rho_p=3000$~kg/m$^3,$  $c_a=896$~J/kg/K, $c_p=500$~J/kg/K,  $\alpha=8.4$~W/m$^2$/K,  $h=0.1$~m,  $r_1=0.031$~m, $z_0=0.00195$~m, $z_1=z_0+h$,  $S=0.0427$~W/K/A,  $R=6.03$~$\Omega,$  $u_+=1.115$~V and $u_-=-1.29$~V.  We take the reference temperature $\theta^0=293$~K, and the initial steady state $\theta_{st}$ such that the average temperature $\theta_{av}$ in the controlled cylinder $\mathcal{V}_1$ is $5.5$~K. The  voltage $u_{st}$ corresponding to this state is found solving (\ref{Minimum Condition Steady State Minus Average}): $u_{st}\approx 1.44$~V. We take this value as the constraints: $|u(t)|\leq 1.44$~V. The terminal time $T$ is varied with the step $\delta T_0=12$~sec.

\begin{figure}
\begin{center}
\includegraphics[width=0.6\textwidth]{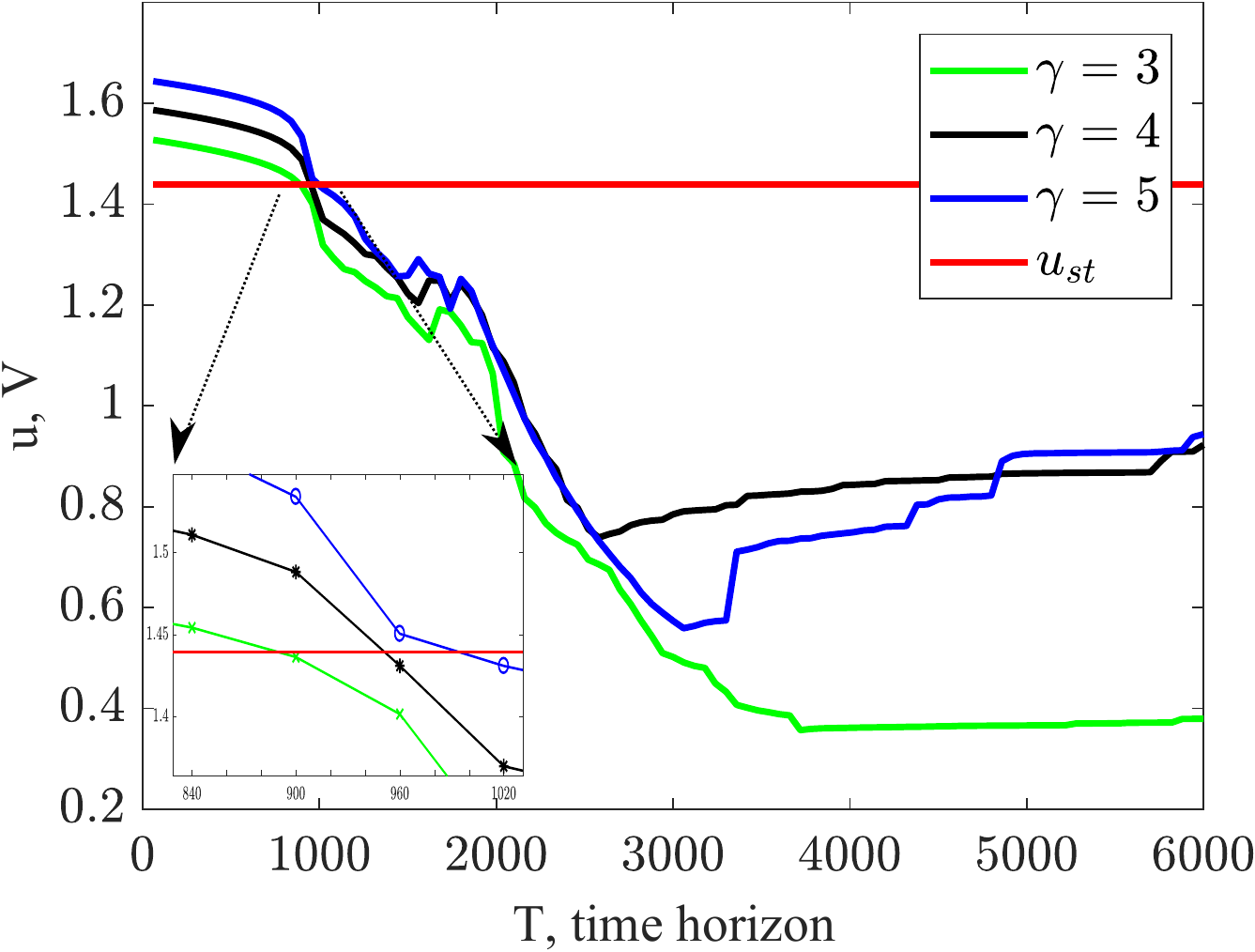}
\caption{The maximum absolute value of the control function $u(t)$: $\max_{t\in[0,T]} |u(t)|$.} 
\label{fig:max u}
\end{center}
\end{figure}

\begin{figure}
\begin{center}
\includegraphics[width=0.6\textwidth]{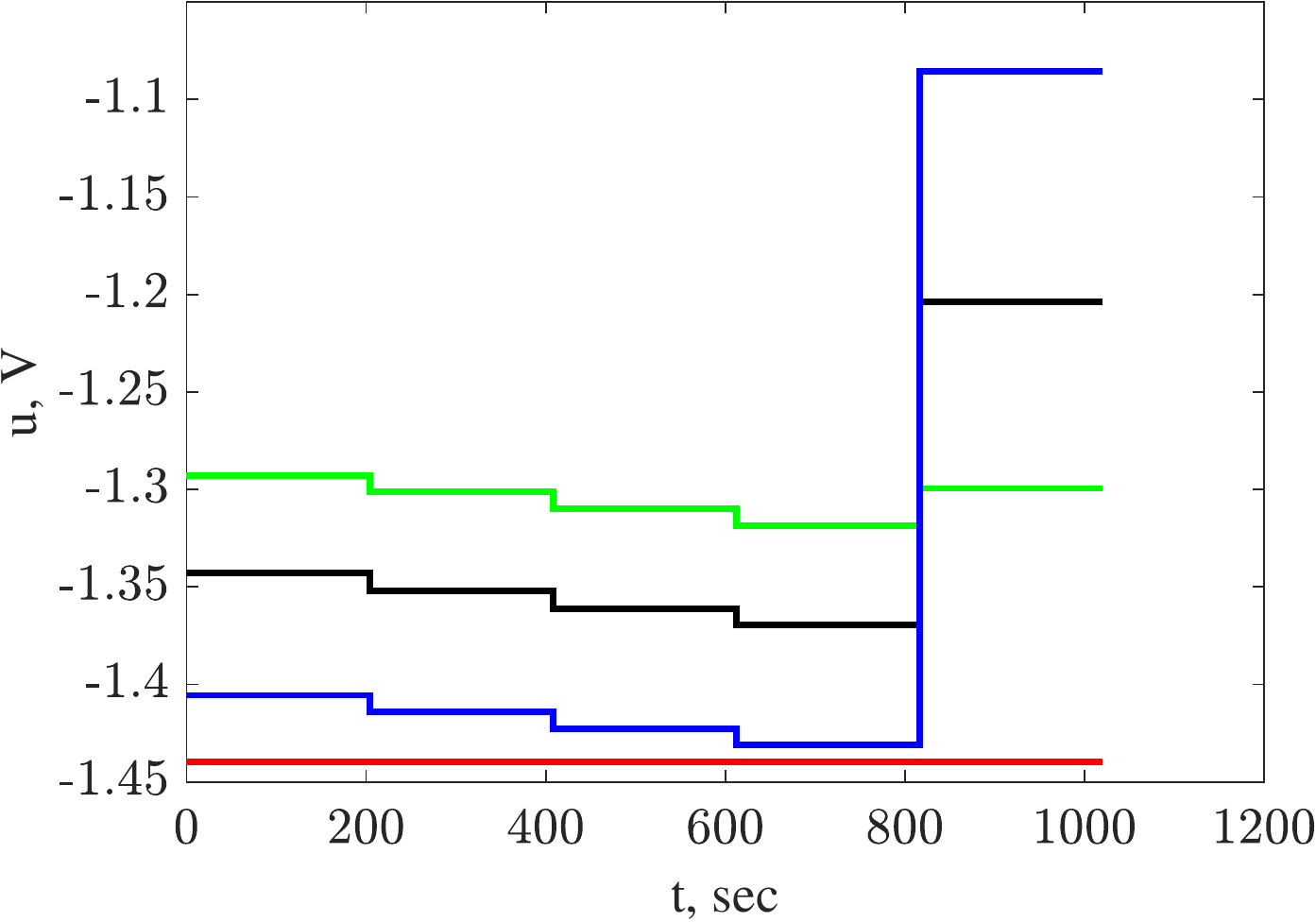}
\caption{The piecewise control function $u(t)$; $T=1020$~sec.} 
\label{fig:u piecewise}
\end{center}
\end{figure}

\begin{figure}
\begin{center}
\includegraphics[width=0.6\textwidth]{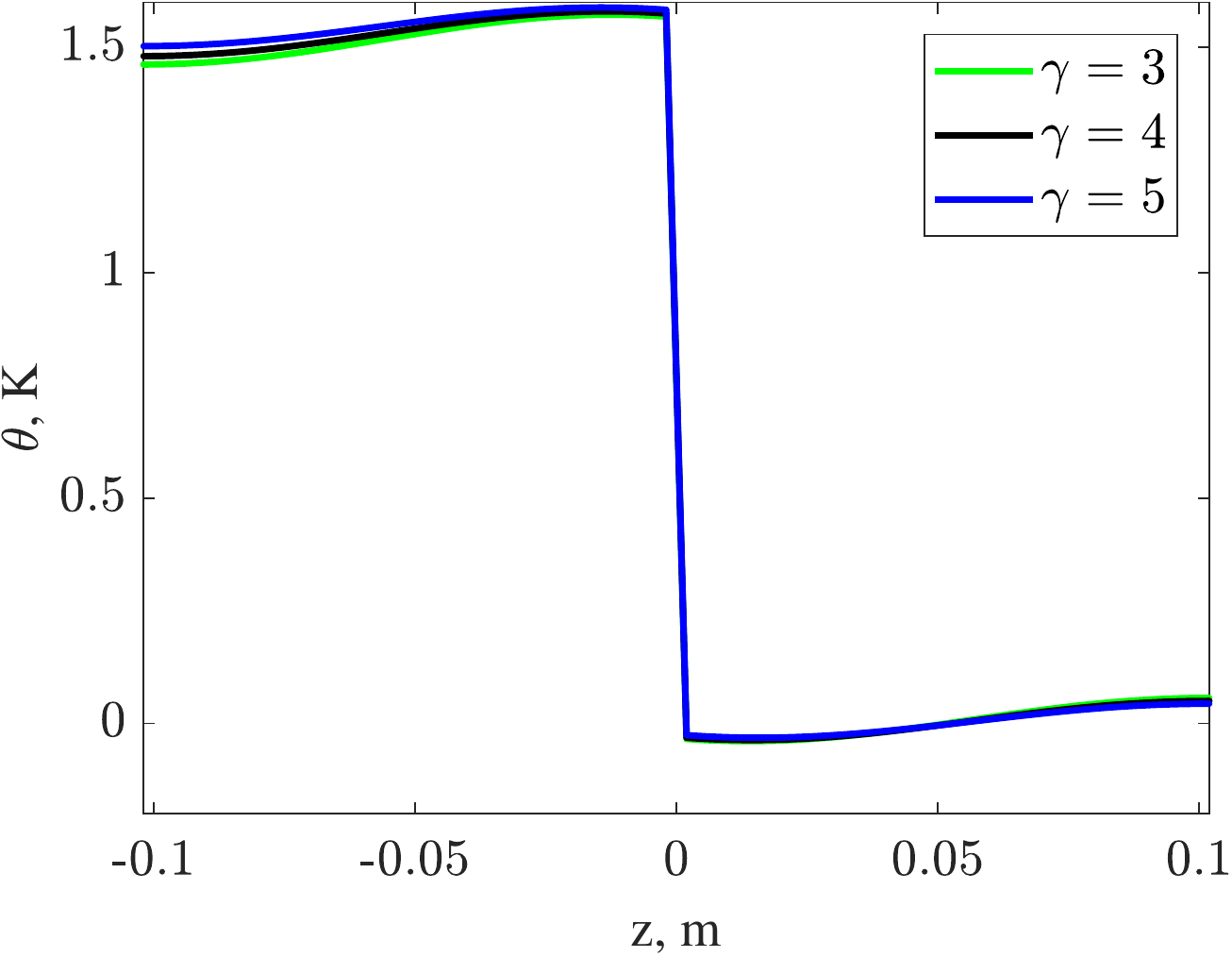}
\caption{The terminal temperature distribution.} 
\label{fig:terminal temperature}
\end{center}
\end{figure}

\begin{figure}
\begin{center}
\includegraphics[width=0.6\textwidth]{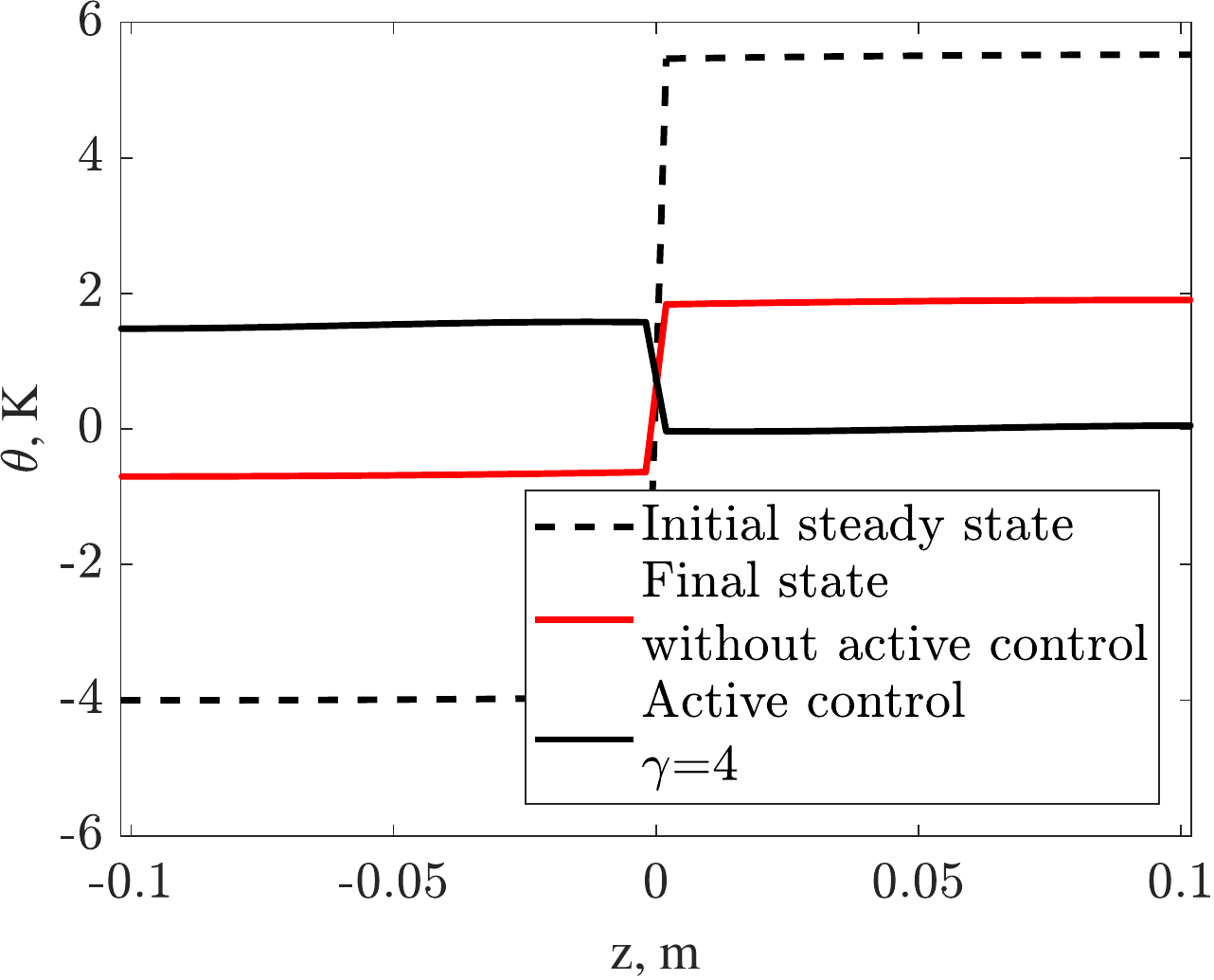}
\caption{The initial and terminal temperature distribution for the controlled and uncontrolled processes.} 
\label{fig:terminal temperature vs natural cooling}
\end{center}
\end{figure}

To design  admissible control laws, we consider several values of the weighting coefficient $\gamma$ in (\ref{Full Cost Functional}). The larger value of $\gamma$ favors minimization of the temperature rather than satisfaction of the control constraints.
In Figs.~\ref{fig:cost functional}--\ref{fig:terminal temperature vs natural cooling} the numerical results for the values of the weighting coefficient $\gamma=3,4,5$ are presented.  In Fig.~\ref{fig:cost functional}, the optimal values of the cost function $F$ (in logarithmic scale) are shown as a function of the time horizon $T$. These values do not change significantly while $T\gg  %T_{min},$ $T_{min}\geq 
10^3$~sec. In this region, the obtained control laws satisfy constraints and transfer the system  to a small neighbourhood of the zero state.

 This statement is supported by Figs.~\ref{fig:temperature normalized by cooling},~\ref{fig:temperature normalized by initial}, where the relative $L_2$-norm of the terminal temperature distribution in $\mathcal{V}_1$ is given in logarithmic scale. In Fig.~\ref{fig:temperature normalized by cooling}, these values are normalized by the norm of the temperature in $\mathcal{V}_1$ that emerges in this cylinder due to natural cooling during the same time $T$. Therefore,   Fig.~\ref{fig:temperature normalized by cooling} shows the effectiveness of the control law. As it can be expected, this effectiveness  grows as $T$ decreases: on long  time intervals the natural cooling is enough to achieve a state close to zero. However, as $T$ becomes smaller, the advantage of an active control is more clear: it allows to make temperature  up to 3 order smaller than the natural cooling. Nevertheless, since  power required to achieve the zero state grows with decreasing $T$, for $T\gtrsim 10^3$ the control law within the constraints loses its advantages and the effectiveness diminishes quickly. Fig.~\ref{fig:temperature normalized by cooling} demonstrates the temperature decreasing comparing with the initial steady state. Here, the $L_2$-norm of the terminal temperature distribution in $\mathcal{V}_1$ (in logarithmic scale) is related to the $L_2$-norm of the steady state. The red curve represents the natural cooling.

Figs.~\ref{fig:max u},~\ref{fig:u piecewise} illustrate the behavior of the optimal control law. Fig.~\ref{fig:max u} shows the maximum absolute value of $u(t)$. This allows us to estimate $T_{min}$. Fix $\varepsilon=10^{-2}.$ Note that relaxing the temperature minimization (making $\gamma$ smaller) let us achieve smaller values of $T_{min}$: $T_{min}$ for $\gamma=3$ is less then $10^3$~sec, but $T_{min}>10^3$~sec for $\gamma=4,5.$ In Fig.~\ref{fig:u piecewise}, the piecewise optimal control laws for $\tilde T=1020$~sec are given. The control constraint is depicted by the red line.  As can be seen for the chosen time step $\delta T_0$, this value $\tilde T$ may be taken as $T_{min}$ for $\gamma=5.$

Figs.~\ref{fig:terminal temperature},~\ref{fig:terminal temperature vs natural cooling} show the terminal temperature distribution for the weighting coefficients $\gamma=3,4,5$. Here, $T=1020$~sec. Although the control laws are rather different, see Fig.~\ref{fig:u piecewise}, these distributions in the controlled cylinder $\mathcal{V}_1$ are close to each other. Note that the distinction between the results of the considered control laws is more pronounced in the uncontrolled cylinder $\mathcal{V}_2$ (negative values of $z$), which is used as a heat sink. In Fig.~~\ref{fig:terminal temperature vs natural cooling}, we may compare the initial steady state (dashed line) with the distributions obtained due to natural cooling (red curve) and due to the optimal control law (black curve) for $\gamma=4$. Note the overheating of the uncontrolled cylinder $\mathcal{V}_2$. The time $T$ is too short to get to the zero state in $\mathcal{V}_1$ via natural cooling and the actuator transfers the energy from $\mathcal{V}_1$ to $\mathcal{V}_2$ by heating it.

\begin{figure}
\begin{center}
\includegraphics[width=0.6\textwidth]{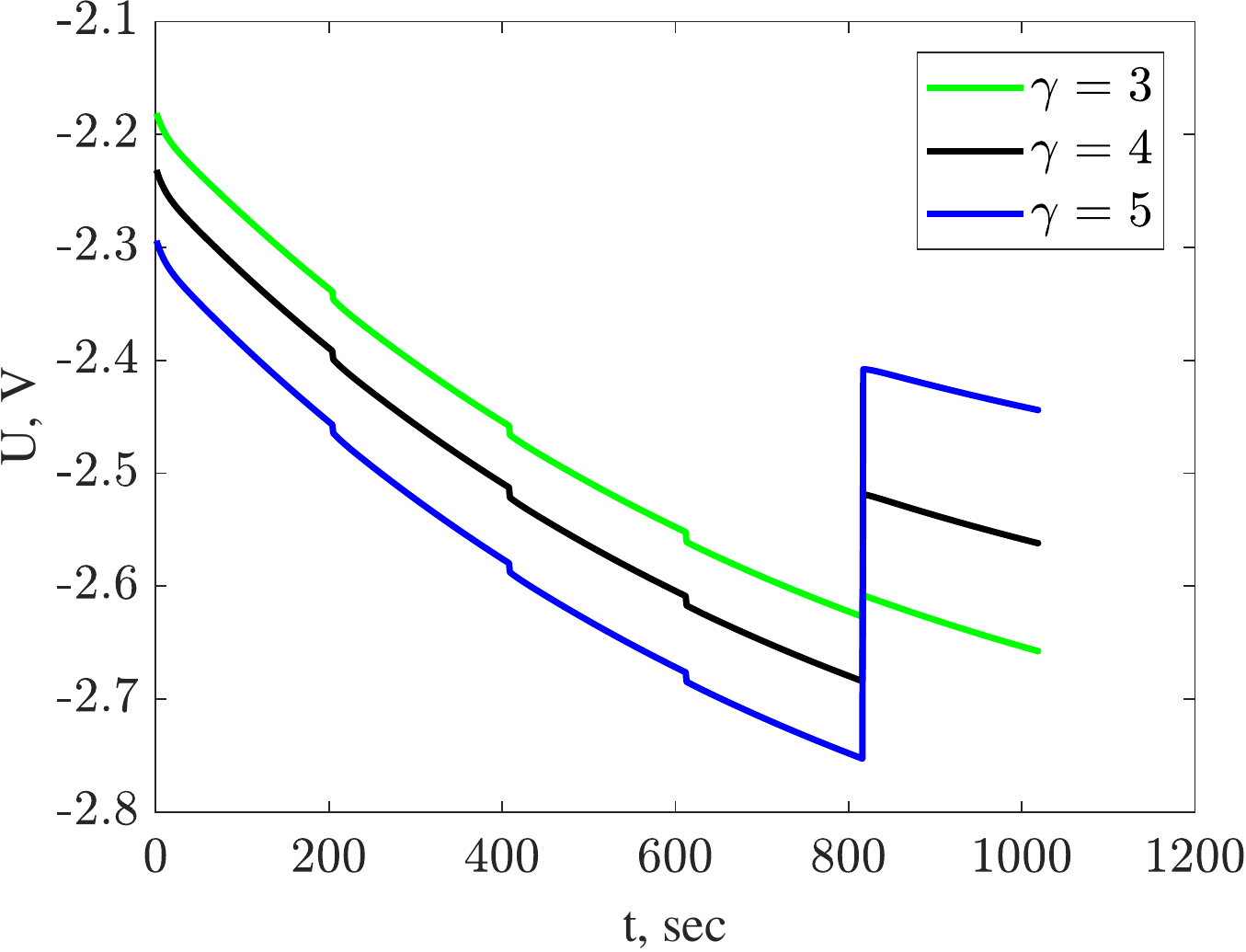}
\caption{Total input signal with feedback.} 
\label{fig:nonlinear behavior}
\end{center}
\end{figure}

Fig.~\ref{fig:nonlinear behavior} demonstrates the nonlinear essence of the studied thermoelectric system. The total input voltage involves both feedforward control law $u(t)$ and the feedback linearization signal $S[\tilde\theta]$ according to (\ref{full control}). Although $u(t)$ is piecewuse constant, the full input voltage $U(t)$ is more sophisticated, cf. Fig.~\ref{fig:u piecewise}. As the electric current in the PE $Ru$ is proportional to the voltage $u$, the term $S[\tilde\theta]$ does not influence on the Joule heat loss. 

\section{Conclusions}
In this paper, we study the constrained time-optimal problem of achieving a prescribed temperature distribution in a thermoelectric solid system. We consider a system consisting of  two identical cylinders with a thin Peltier element between them. Although the entire system is actuated by the thermoelectric converter, we try to reach the desired state only in one of them, treating the other cylinder as a heat sink.  Assuming that the optimal control law is sought in the class of piecewise constant functions, we reduce the nonlinear PDE system to its finite-dimensional approximation by utilizing eigenfunction decomposition. The constraints are implemented via a penalty term in the cost functional. Next, we vary the time horizon  finding an optimal control law with applying the gradient descend method at each step of the variation. On each iteration of the gradient descend, the direct problem is solved explicitly. It has been shown that this combined numerical-analytical approach allows us to reach a rather small neighborhood of the prescribed state without violation of the constraints and estimate from above the minimal control time. We plan to verify proposed strategy and the finite-dimensional approximation exploited by implementing the designed control laws in a FEM setting. Next, we aim at controlling a structure consisting of several heat conducting bodies with Peltier elements between them.

\end{document}